
\font\twelverm=cmr10 scaled 1200    \font\twelvei=cmmi10 scaled 1200
\font\twelvesy=cmsy10 scaled 1200   \font\twelveex=cmex10 scaled 1200
\font\twelvebf=cmbx10 scaled 1200   \font\twelvesl=cmsl10 scaled 1200
\font\twelvett=cmtt10 scaled 1200   \font\twelveit=cmti10 scaled 1200
\skewchar\twelvei='177   \skewchar\twelvesy='60
\def\twelvepoint{\normalbaselineskip=12.4pt
  \abovedisplayskip 12.4pt plus 3pt minus 9pt
  \belowdisplayskip 12.4pt plus 3pt minus 9pt
  \abovedisplayshortskip 0pt plus 3pt
  \belowdisplayshortskip 7.2pt plus 3pt minus 4pt
  \smallskipamount=3.6pt plus1.2pt minus1.2pt
  \medskipamount=7.2pt plus2.4pt minus2.4pt
  \bigskipamount=14.4pt plus4.8pt minus4.8pt
  \def\rm{\fam0\twelverm}          \def\it{\fam\itfam\twelveit}%
  \def\sl{\fam\slfam\twelvesl}     \def\bf{\fam\bffam\twelvebf}%
  \def\mit{\fam 1}                 \def\cal{\fam 2}%
  \def\tt{\twelvett}
  \textfont0=\twelverm   \scriptfont0=\tenrm   \scriptscriptfont0=\sevenrm
  \textfont1=\twelvei    \scriptfont1=\teni    \scriptscriptfont1=\seveni
  \textfont2=\twelvesy   \scriptfont2=\tensy   \scriptscriptfont2=\sevensy
  \textfont3=\twelveex   \scriptfont3=\twelveex  \scriptscriptfont3=\twelveex
  \textfont\itfam=\twelveit
  \textfont\slfam=\twelvesl
  \textfont\bffam=\twelvebf \scriptfont\bffam=\tenbf
  \scriptscriptfont\bffam=\sevenbf
  \normalbaselines\rm}

\def\beginlinemode{\endmode
  \begingroup\parskip=0pt \obeylines\def\\{\par}\def\endmode{\par\endgroup}}
\def\beginparmode{\endmode
  \begingroup \def\endmode{\par\endgroup}}
\let\endmode=\par
{\obeylines\gdef\
{}}
\def\singlespace{\baselineskip=\normalbaselineskip}
\def\oneandthreefifthsspace{\baselineskip=\normalbaselineskip
  \multiply\baselineskip by 8 \divide\baselineskip by 5}

\def\oneandahalfspace{\baselineskip=\normalbaselineskip
  \multiply\baselineskip by 3 \divide\baselineskip by 2}
\def\doublespace{\baselineskip=\normalbaselineskip \multiply\baselineskip by 2}
\newcount\firstpageno
\firstpageno=2
\footline={
\ifnum\pageno<\firstpageno{\hfil}\else{\hfil\twelverm\folio\hfil}\fi}
\let\rawfootnote=\footnote              
\def\footnote#1#2{{\rm\singlespace\parindent=0pt\rawfootnote{#1}{#2}}}
\def\raggedcenter{\leftskip=2em plus 12em \rightskip=\leftskip
  \parindent=0pt \parfillskip=0pt \spaceskip=.3333em \xspaceskip=.5em
  \pretolerance=9999 \tolerance=9999
  \hyphenpenalty=9999 \exhyphenpenalty=9999 }
\def\veryraggedcenter{\leftskip=.1em plus 6em \rightskip=\leftskip
  \parindent=0pt \parfillskip=0pt \spaceskip=.3333em \xspaceskip=.5em
  \pretolerance=9999 \tolerance=9999
  \hyphenpenalty=9999 \exhyphenpenalty=9999 }
\parskip=\medskipamount
\twelvepoint            
\overfullrule=0pt       
\def\preprintno#1{
 \rightline{\rm #1}}    
\def\author                     
  {\vskip 3pt plus 0.2fill \beginlinemode
   \singlespace \veryraggedcenter \twelvesc}
\def\affil                      
  {\vskip 3pt plus 0.1fill \beginlinemode
   \oneandahalfspace \raggedcenter \sl}
\def\abstract                   
  {\vskip 3pt plus 0.3fill \beginparmode
   \doublespace \narrower \noindent ABSTRACT: }
\def\abstract                   
  {\vskip 3pt plus 0.3fill \beginparmode
   \oneandthreefifthsspace \narrower \noindent ABSTRACT: }
\def\endtitlepage               
  {\endpage                     
   \body}
\def\body                       
  {\beginparmode}               

\def\subhead#1{                 
  \vskip 0.25truein             
  {\raggedcenter #1 \par}
   \nobreak\vskip 0.1truein\nobreak}
\def\refto#1{$|{#1}$}           
\def\references                 
  {\subhead{References}         
   \beginparmode
   \frenchspacing \parindent=0pt \leftskip=1truecm
   \parskip=8pt plus 3pt \everypar{\hangindent=\parindent}}
\gdef\refis#1{\indent\hbox to 0pt{\hss#1.~}}    
\gdef\journal#1, #2, #3, 1#4#5#6{               
    {\sl #1~}{\bf #2}, #3, (1#4#5#6)}           
\def\refstylenp{                
  \gdef\refto##1{ [##1]}                                
  \gdef\refis##1{\indent\hbox to 0pt{\hss##1)~}}        
  \gdef\journal##1, ##2, ##3, ##4 {                     
     {\sl ##1~}{\bf ##2~}(##3) ##4 }}
\def\refstyleprnp{              
  \gdef\refto##1{ [##1]}                                
  \gdef\refis##1{\indent\hbox to 0pt{\hss##1)~}}        
  \gdef\journal##1, ##2, ##3, 1##4##5##6{               
    {\sl ##1~}{\bf ##2~}(1##4##5##6) ##3}}
\def\pr{\journal Phys. Rev., }

\def\prl{\journal Phys. Rev. Lett., }
\def\prpts{\journal Phys. Rep., }
\def\np{\journal Nucl. Phys., }
\def\pl{\journal Phys. Lett., }

\def\endreferences{\body}
\def\endpage                    
  {\vfill\eject}
\def\endpaper                   
  {\endmode\vfill\supereject}
\def\endit
  {\endpaper\end}
\def\ref#1{Ref. #1}                     
\def\Ref#1{Ref. #1}                     

\def\m@th{\mathsurround=0pt }
\font\twelvesc=cmcsc10 scaled 1200
\def\cite#1{{#1}}
\def\(#1){(\call{#1})}
\def\call#1{{#1}}
\def\taghead#1{}
\def\leaderfill{\leaders\hbox to 1em{\hss.\hss}\hfill}
\def\twiddle{\lower.9ex\rlap{$\kern-.1em\scriptstyle\sim$}}
\def\bigtwiddle{\lower1.ex\rlap{$\sim$}}
\def\gtwid{\mathrel{\raise.3ex\hbox{$>$\kern-.75em\lower1ex\hbox{$\sim$}}}}
\def\ltwid{\mathrel{\raise.3ex\hbox{$<$\kern-.75em\lower1ex\hbox{$\sim$}}}}
\def\square{\kern1pt\vbox{\hrule height 1.2pt\hbox{\vrule width 1.2pt\hskip 3pt
   \vbox{\vskip 6pt}\hskip 3pt\vrule width 0.6pt}\hrule height 0.6pt}\kern1pt}
\catcode`@=11
\newcount\tagnumber\tagnumber=0

\immediate\newwrite\eqnfile
\newif\if@qnfile\@qnfilefalse
\def\write@qn#1{}
\def\writenew@qn#1{}
\def\w@rnwrite#1{\write@qn{#1}\message{#1}}
\def\@rrwrite#1{\write@qn{#1}\errmessage{#1}}

\def\taghead#1{\gdef\t@ghead{#1}\global\tagnumber=0}
\def\t@ghead{}

\expandafter\def\csname @qnnum-3\endcsname
  {{\t@ghead\advance\tagnumber by -3\relax\number\tagnumber}}
\expandafter\def\csname @qnnum-2\endcsname
  {{\t@ghead\advance\tagnumber by -2\relax\number\tagnumber}}
\expandafter\def\csname @qnnum-1\endcsname
  {{\t@ghead\advance\tagnumber by -1\relax\number\tagnumber}}
\expandafter\def\csname @qnnum0\endcsname
  {\t@ghead\number\tagnumber}
\expandafter\def\csname @qnnum+1\endcsname
  {{\t@ghead\advance\tagnumber by 1\relax\number\tagnumber}}
\expandafter\def\csname @qnnum+2\endcsname
  {{\t@ghead\advance\tagnumber by 2\relax\number\tagnumber}}
\expandafter\def\csname @qnnum+3\endcsname
  {{\t@ghead\advance\tagnumber by 3\relax\number\tagnumber}}

\def\equationfile{%
  \@qnfiletrue\immediate\openout\eqnfile=\jobname.eqn%
  \def\write@qn##1{\if@qnfile\immediate\write\eqnfile{##1}\fi}
  \def\writenew@qn##1{\if@qnfile\immediate\write\eqnfile
    {\noexpand\tag{##1} = (\t@ghead\number\tagnumber)}\fi}
}
\def\callall#1{\xdef#1##1{#1{\noexpand\call{##1}}}}
\def\call#1{\each@rg\callr@nge{#1}}
\def\each@rg#1#2{{\let\thecsname=#1\expandafter\first@rg#2,\end,}}
\def\first@rg#1,{\thecsname{#1}\apply@rg}
\def\apply@rg#1,{\ifx\end#1\let\next=\relax%
\else,\thecsname{#1}\let\next=\apply@rg\fi\next}
\def\callr@nge#1{\calldor@nge#1-\end-}
\def\callr@ngeat#1\end-{#1}
\def\calldor@nge#1-#2-{\ifx\end#2\@qneatspace#1 %
  \else\calll@@p{#1}{#2}\callr@ngeat\fi}
\def\calll@@p#1#2{\ifnum#1>#2{\@rrwrite{Equation range #1-#2\space is bad.}
\errhelp{If you call a series of equations by the notation M-N, then M and
N must be integers, and N must be greater than or equal to M.}}\else%
 {\count0=#1\count1=#2\advance\count1
by1\relax\expandafter\@qncall\the\count0,%
  \loop\advance\count0 by1\relax%
    \ifnum\count0<\count1,\expandafter\@qncall\the\count0,%
  \repeat}\fi}

\def\@qneatspace#1#2 {\@qncall#1#2,}
\def\@qncall#1,{\ifunc@lled{#1}{\def\next{#1}\ifx\next\empty\else
  \w@rnwrite{Equation number \noexpand\(>>#1<<) has not been defined yet.}
  >>#1<<\fi}\else\csname @qnnum#1\endcsname\fi}
\let\eqnono=\eqno
\def\eqno(#1){\tag#1}
\def\tag#1$${\eqnono(\displayt@g#1 )$$}
\def\aligntag#1\endaligntag
  $${\gdef\tag##1\\{&(##1 )\cr}\eqalignno{#1\\}$$
  \gdef\tag##1$${\eqnono(\displayt@g##1 )$$}}

\def\eqalignno#1{\displ@y \tabskip\centering
  \halign to\displaywidth{\hfil$\displaystyle{##}$\tabskip\z@skip
    &$\displaystyle{{}##}$\hfil\tabskip\centering
    &\llap{$\displayt@gpar##$}\tabskip\z@skip\crcr
    #1\crcr}}

\def\displayt@gpar(#1){(\displayt@g#1 )}

\def\displayt@g#1 {\rm\ifunc@lled{#1}\global\advance\tagnumber by1
        {\def\next{#1}\ifx\next\empty\else\expandafter
        \xdef\csname @qnnum#1\endcsname{\t@ghead\number\tagnumber}\fi}%
  \writenew@qn{#1}\t@ghead\number\tagnumber\else
        {\edef\next{\t@ghead\number\tagnumber}%
        \expandafter\ifx\csname @qnnum#1\endcsname\next\else
        \w@rnwrite{Equation \noexpand\tag{#1} is a duplicate number.}\fi}%
  \csname @qnnum#1\endcsname\fi}

\def\ifunc@lled#1{\expandafter\ifx\csname @qnnum#1\endcsname\relax}

\let\@qnend=\end\gdef\end{\if@qnfile
\immediate\write16{Equation numbers written on []\jobname.EQN.}\fi\@qnend}

\catcode`@=12
\refstyleprnp
\catcode`@=11
\newcount\r@fcount \r@fcount=0
\def\refreset{\global\r@fcount=0}
\newcount\r@fcurr
\immediate\newwrite\reffile
\newif\ifr@ffile\r@ffilefalse
\def\w@rnwrite#1{\ifr@ffile\immediate\write\reffile{#1}\fi\message{#1}}

\def\writer@f#1>>{}
\def\referencefile{
  \r@ffiletrue\immediate\openout\reffile=\jobname.ref%
  \def\writer@f##1>>{\ifr@ffile\immediate\write\reffile%
    {\noexpand\refis{##1} = \csname r@fnum##1\endcsname = %
     \expandafter\expandafter\expandafter\strip@t\expandafter%
     \meaning\csname r@ftext\csname r@fnum##1\endcsname\endcsname}\fi}%
  \def\strip@t##1>>{}}

\def\citeall#1{\xdef#1##1{#1{\noexpand\cite{##1}}}}
\def\cite#1{\each@rg\citer@nge{#1}}	

\def\each@rg#1#2{{\let\thecsname=#1\expandafter\first@rg#2,\end,}}
\def\first@rg#1,{\thecsname{#1}\apply@rg}	
\def\apply@rg#1,{\ifx\end#1\let\next=\relax
\else,\thecsname{#1}\let\next=\apply@rg\fi\next}

\def\citer@nge#1{\citedor@nge#1-\end-}	
\def\citer@ngeat#1\end-{#1}
\def\citedor@nge#1-#2-{\ifx\end#2\r@featspace#1 
  \else\citel@@p{#1}{#2}\citer@ngeat\fi}	
\def\citel@@p#1#2{\ifnum#1>#2{\errmessage{Reference range #1-#2\space is bad.}%
    \errhelp{If you cite a series of references by the notation M-N, then M and
    N must be integers, and N must be greater than or equal to M.}}\else%
 {\count0=#1\count1=#2\advance\count1
by1\relax\expandafter\r@fcite\the\count0,%
  \loop\advance\count0 by1\relax
    \ifnum\count0<\count1,\expandafter\r@fcite\the\count0,%
  \repeat}\fi}

\def\r@featspace#1#2 {\r@fcite#1#2,}	
\def\r@fcite#1,{\ifuncit@d{#1}
    \newr@f{#1}%
    \expandafter\gdef\csname r@ftext\number\r@fcount\endcsname%
                     {\message{Reference #1 to be supplied.}%
                      \writer@f#1>>#1 to be supplied.\par}%
 \fi%
 \csname r@fnum#1\endcsname}
\def\ifuncit@d#1{\expandafter\ifx\csname r@fnum#1\endcsname\relax}%
\def\newr@f#1{\global\advance\r@fcount by1%
    \expandafter\xdef\csname r@fnum#1\endcsname{\number\r@fcount}}

\let\r@fis=\refis			
\def\refis#1#2#3\par{\ifuncit@d{#1}
   \newr@f{#1}%
   \w@rnwrite{Reference #1=\number\r@fcount\space is not cited up to now.}\fi%
  \expandafter\gdef\csname r@ftext\csname r@fnum#1\endcsname\endcsname%
  {\writer@f#1>>#2#3\par}}

\def\ignoreuncited{
   \def\refis##1##2##3\par{\ifuncit@d{##1}%
     \else\expandafter\gdef\csname r@ftext\csname
r@fnum##1\endcsname\endcsname%
     {\writer@f##1>>##2##3\par}\fi}}

\def\r@ferr{\endreferences\errmessage{I was expecting to see
\noexpand\endreferences before now;  I have inserted it here.}}
\let\r@ferences=\references
\def\references{\r@ferences\def\endmode{\r@ferr\par\endgroup}}

\let\endr@ferences=\endreferences
\def\endreferences{\r@fcurr=0
  {\loop\ifnum\r@fcurr<\r@fcount
    \advance\r@fcurr by 1\relax\expandafter\r@fis\expandafter{\number\r@fcurr}%
    \csname r@ftext\number\r@fcurr\endcsname%
  \repeat}\gdef\r@ferr{}\global\r@fcount=0\endr@ferences}

\let\r@fend=\endpaper\gdef\endpaper{\ifr@ffile
\immediate\write16{Cross References written on []\jobname.REF.}\fi\r@fend}

\catcode`@=12

\citeall\refto		
\citeall\ref		%
\citeall\Ref		%

\referencefile
\def\h{h^0}
\def\mh{m_{h^0}}
\def\mhsquared{m^2_{h^0}}
\def\mhcubed{m^3_{h^0}}

\def\umichadd{Randall Physics Laboratory\\University of Michigan
\\Ann Arbor MI 48109-1120}

\def\oneandthreefifthsspace{\baselineskip=\normalbaselineskip
  \multiply\baselineskip by 8 \divide\baselineskip by 5}

\font\titlefont=cmr10 scaled\magstep3
\def\bigtitle                      
  {\null\vskip 3pt plus 0.2fill
   \beginlinemode \doublespace \raggedcenter \titlefont}

\def\lsp{\tilde\chi_1^0}
\def\mlsp{m_{\tilde\chi_1^0}}

\def\lsp{\tilde N_1}
\def\mlsp{m_{\tilde N_1}}

\hyphenation{Nil-les}

\oneandthreefifthsspace
\preprintno{hep-ph/9508265}
\bigtitle{Two-photon decays of the lightest Higgs
            boson of supersymmetry at the LHC}
\medskip
\author
G.~L.~Kane, Graham~D.~Kribs, Stephen~P.~Martin, and James~D.~Wells$^\dagger$
\affil\umichadd
\body
\footnote{}{$^\dagger$ Address after Sept.~1, 1995: SLAC, MS 81,
P.O.~Box 4349, Stanford CA 94309}

\abstract
We discuss the production and two-photon decay of the lightest Higgs boson
($\h $) of the minimal supersymmetric standard model at the CERN Large Hadron
Collider. Since the observability of the signal is quite model dependent, we
conduct a thorough scan of the parameter space of minimal supersymmetry,
including experimental and theoretical constraints. If kinematically allowed,
supersymmetric decay modes of $\h $ may be important, and can even dominate
all others. The coupling of $\h $ to $b\overline b$ can be different from that
of a standard model Higgs boson; this can diminish (or enhance, but only if
$\tan\beta$ is very large) the $\h \rightarrow\gamma\gamma $ signal. We
emphasize the importance of a full treatment of radiative corrections in the
Higgs sector for obtaining the $\h b\overline b$ coupling. If supersymmetric
particles are not too heavy, their contributions in loops can either enhance
or suppress both the production cross-section and the $\h \rightarrow
\gamma\gamma$ branching fraction. We discuss the relative importance of these
factors in the context of various scenarios for the discovery of supersymmetry.
Even if $\h$ is not detected at the LHC, $\h$ may still
exist in its expected mass region.

\endtitlepage
\oneandthreefifthsspace

One of the main reasons for the construction of new high-energy colliders is
to find the Higgs scalar boson(s) associated with electroweak symmetry breaking
in the standard model and its perturbative extensions. In a completely
general framework of electroweak symmetry breaking, the Higgs sector is only
weakly constrained. However, softly-broken supersymmetry is the only
known theory in which a fundamental Higgs scalar does not receive
disastrously large radiative corrections from physics at arbitrarily
high energy scales.
This theoretical perspective encourages and perhaps demands
the presumption that nature is
supersymmetric if a fundamental Higgs boson exists at all.

Low-energy supersymmetry does imply non-trivial constraints
on the Higgs sector.
The minimal supersymmetric standard model (MSSM) [\cite{reviews}]
contains two Higgs doublet
chiral superfields, which after electroweak symmetry breaking result in
two CP-even neutral scalars (conventionally denoted $\h $ and $H^0$),
one CP-odd
neutral scalar ($A^0$) and a pair of charged scalars ($H^\pm$). The masses of
the heavier CP-even neutral scalar $H^0$ and of $A^0, H^\pm$ are
constrained only weakly by subjective criteria such as fine-tuning
[\cite{finetuning}]; they therefore might escape detection at all colliders
currently being planned.
Fortunately, the mass of the lighter CP-even Higgs boson $\h $ is guaranteed
to be less than about 140 GeV in the MSSM, and less than about 150 GeV
in other supersymmetric models which remain perturbative
up to very high energies [\cite{kkw}].
This means that $\h $ is certainly kinematically accessible to future colliders
and should eventually be discovered at a high energy $e^+e^-$ collider (NLC)
with $\sqrt s \geq 250$ GeV if it exists.
Of course one would like to detect the Higgs sector before
an NLC is built, but if the mass of $\h $ exceeds the mass of the $Z$ boson by
more than a few GeV, it will probably escape detection at LEP-II.
If the Tevatron
collects $\geq 30$ fb$^{-1}$ of data it has been argued [\cite{fnal}]
that a Higgs boson is also detectable there up to and perhaps above 130 GeV,
if it decays with similar branching fractions as the standard model Higgs.

In this paper we will examine the possibility of detecting $\h$ produced
in $pp$ collisions at the CERN Large Hadron Collider (LHC) operating
at $\sqrt{s} = 14$ TeV.
The most important production mechanism for $\h$ is by gluon fusion
through quark- and squark-loop graphs.
The tree-level decays of $\h \rightarrow b\overline b$ will
dominate (unless decays
of $\h$ into pairs of supersymmetric particles are kinematically allowed,
as we shall see), but are probably
not useful because of large hadronic backgrounds.
Instead, the detection of $\h$ relies on the rare decay $\h \rightarrow
\gamma\gamma$, which proceeds via one-loop graphs with all possible charged
particles running in the loop. This strategy requires excellent energy
resolution (on the order of 1\%) for the $\gamma\gamma$ pairs
as well as good jet rejection to overcome the standard model backgrounds, which
come from $q\overline q\rightarrow \gamma\gamma$ and $gg\rightarrow
\gamma\gamma$ [\cite{backgrounds}] as well as jets imitating a $\gamma$
in the detector.
\phantom{\cite{ehsv}\cite{gkw}\cite{ggn}\cite{bbsp}\cite{bbkt92}\cite{bbdkt}
\cite{kz}\cite{cms}\cite{atlas}\cite{sdgz}\cite{kop}\cite{konig}
}

Many groups have studied [\cite{ehsv}-\cite{konig}]
the viability of this signal (as well as $H^0\rightarrow \gamma\gamma$ and
$A^0\rightarrow \gamma\gamma$)
at proton-proton supercolliders. However, most of these studies have
employed special choices for the MSSM model parameters, or neglected the
possibility of supersymmetric decays for $\h$, so that it is difficult to
gain an understanding of how general the results are.
The total cross-section times branching ratio
for $ pp \rightarrow \h \rightarrow \gamma\gamma $ can depend
quite strongly on the masses and couplings of the superpartners and Higgs
bosons, particularly if they are not too heavy.
Therefore, it is of utmost importance to consider all possibilities for the
unknown model parameters. In this paper we will consider the relevant
production cross section, decay widths and branching ratios
for an ensemble of models within the
general framework of the MSSM, each chosen to satisfy all present
experimental constraints.

A completely general supersymmetrization of the standard model would
introduce more than a hundred new parameters associated with the soft
supersymmetry-breaking masses and couplings of the superpartners. Fortunately,
we already have experimental evidence that these parameters are far from
arbitrary; otherwise, large flavor-changing neutral currents would be expected
to manifest themselves in processes such as
$K\leftrightarrow\overline K$ mixing, $b\rightarrow s \gamma$, and
$\mu \rightarrow e \gamma$. Thus there is strong circumstantial evidence
in favor of some ``organizing principle" governing
the MSSM model parameters. On the theoretical side, supergravity models
[\cite{sugra}] provide just such an organizing principle, so that the
features of the model may depend on only a few parameters at the scale
$M_X \approx 2 \times 10^{16}$ GeV where low-energy data indicates a
unification of gauge couplings. This ``super-unified" model framework may have
to be modified to account for high-energy threshold effects and
other perturbations, but it is both sufficiently
flexible and compelling to support the expectation that it will share
its most important features with the correct theory.

To be concrete, for most of our analysis we generate models randomly
within the super-unified MSSM framework, incorporating radiative electroweak
breaking and gauge coupling unification and universal soft
supersymmetry-breaking terms at $M_X$, using a computer program
similar to the one described
in [\cite{kkrw}]. We take our input parameters to lie in the ranges
$$
\eqalign{
160 \> {\rm GeV} \> < \> &m_{\rm top} < 190 \> {\rm GeV} \cr
1 \> < \> & \tan \beta \> < \> 60  \cr
}\qquad\>\>\>
\eqalign{
0 \> {\rm GeV}\>  < \> &m_0 \> < \> 1000 \> {\rm GeV} \cr
40 \> {\rm GeV} \> < \> &m_{1/2} < \> 500 \> {\rm GeV} \> . \cr
}\eqno(ranges)
$$
The allowed range of the scalar trilinear coupling parameter $A_0$ at $M_X$
is determined by the constraint that there be no charge- and color-breaking
global minima of the scalar potential at the electroweak scale.
We search for such minima in the $SU(3)_C$ D-flat directions
proportional to
$(H_u, \tilde t_L, \tilde t_R, \tilde \nu_\tau) \sim (1,a,a,b)$ and
$(H_d, \tilde b_L, \tilde b_R, \tilde\tau_L,\tilde\tau_R )
\sim (1,c,c,d,d)$, with $a,b,c,d$
arbitrary, using methods indicated in [\cite{clm}]. The resulting constraints
subsume the more traditional but less powerful ones
which correspond to $a=1,b=0$; $c=1,d=0$; and $c=0,d=1$.

On this parameter space we impose experimental constraints, including
direct and indirect sparticle and Higgs boson mass limits [\cite{pdg}];
the invisible width of the Z constraint;
a lightest supersymmetric particle (LSP) which is the lightest neutralino;
a predicted branching ratio for $b\rightarrow s \gamma$ which is
not in gross conflict with recent experimental data [\cite{bsgamma}]; and
the constraint that the relic density of dark matter in the form of
LSPs is not large enough to overclose the universe [\cite{dn}].
(We assume that R-parity is exactly conserved so that the LSP is absolutely
stable.) The resulting constrained parameter space has many correlations
between the masses and couplings of the superpartners and Higgs bosons, not
all of which are immediately obvious in any analytic form.

We compute the Higgs masses and couplings incorporating one-loop
radiative corrections using the effective potential method [\cite{effpot}].
This is an important aspect of the analysis since
large squark mixing and/or nondegeneracy
can induce large corrections to $\mh$ and to the Higgs mixing angle $\alpha$,
and so must not be ignored. Since it will be of
particular importance in the discussion to follow, we pause to make some
remarks on this point.
The (mass)$^2$ matrix for the CP-even neutral scalars $\h,H^0$ is given
by
$$
{\cal M}^2 =
\pmatrix{\sin^2\beta M_{A^0}^2 + \cos^2\beta M_Z^2
& -\sin\beta\cos\beta (M_{A^0}^2 + M_Z^2) \cr
-\sin\beta\cos\beta (M_{A^0}^2 + M_Z^2)
& \cos^2\beta M_{A^0}^2 + \sin^2\beta M_Z^2\cr}
+ \pmatrix{ \Delta_{11} & \Delta_{12} \cr
            \Delta_{12} & \Delta_{22} \cr }
\eqno(hHmatrix)
$$
where the $\Delta_{ij}$ represent radiative corrections which may be found
explicitly e.g.~in [\cite{effpot}]. Then $m_{h^0}^2$
and $m_{H^0}^2$ are the eigenvalues of
\(hHmatrix), and the Higgs mixing angle $\alpha$
is determined by the orthogonal matrix
$$
\pmatrix{ \cos\alpha & \sin\alpha \cr -\sin\alpha & \cos\alpha \cr}
$$
which diagonalizes \(hHmatrix). In the super-unified MSSM parameter space,
one must have $\pi/4 < \beta < \pi/2$ in order to maintain the
perturbativity of the Yukawa couplings up to $M_X$. There
is always a strong correlation between $\alpha$ and $\beta$ if $\tan\beta$ is
not too large, and if $m_{A^0}^2$ dominates over $M_Z^2$ and the radiative
corrections $\Delta_{ij}$.
In this limit one will always have $-\pi/4 < \alpha < 0$, and in particular
$\alpha \approx \beta - \pi/2$. The couplings of $\h$
to electroweak vector bosons $V$ and to fermions are then quite close
to the same couplings for a standard model Higgs boson $\phi^0$. The
ratio of the $\h V V$ couplings to the $\phi^0 V V$ ones are given by
$\sin(\beta - \alpha)$. Within our constrained parameter space,
$\sin(\beta - \alpha )$ is always very close to 1 except when $\tan\beta > 30$
and ${A^0}$ is light. Nevertheless, the ratio of the $\h b\overline b$ coupling
to that of $\phi^0 b \overline b$, given by $-\sin\alpha/\cos\beta$,
can be significantly greater than 1 for any value of $\tan\beta$.
However, we find that $-\sin \alpha/\cos\beta $ cannot be
significantly less than 1 in models with
$\tan\beta < 30$. For $\tan\beta > 30$, $-\sin\alpha/\cos\beta$ is a more
volatile function of the radiative corrections, because
the contributions to the
off-diagonal elements of ${\cal M}^2$ which are proportional to
$m_{A^0}^2$ are highly suppressed, and no longer dominate over
even moderately-sized  $\Delta_{12}$. Thus
one can and does find perfectly viable models with $\tan\beta > 30$ and
$-\sin\alpha/\cos\beta \ll 1$, and even with $\alpha > 0$.
Note that these possibilities will necessarily be missed in commonly-used
treatments which neglect squark mixing and
approximate all squark masses with a single supersymmetry scale  (thus
expressing all radiative corrections to the Higgs sector in terms of
a single parameter $\epsilon$), because
these approximations are tantamount to setting $\Delta_{12}$ to 0 by hand.

Now, the large $\tan\beta$ case might be regarded as merely an
academic curiosity, because as is well known it seems to require some
fine-tuning of the soft parameters in order to achieve
correct electroweak symmetry breaking. While we are sympathetic to this
outlook, it is also true that certain theoretical frameworks (e.g.~those
with Yukawa coupling unification such as $SO(10)$ models) favor a very large
$\tan\beta$. So, for both theoretical and phenomenolgical reasons,
we find it appropriate to divide those parts of the analysis below which
directly involve the $\h b\overline b$ coupling into regions of
high and low $\tan\beta$, with $\tan\beta = 30$ an empirically good
dividing line.

A significant portion of the constrained MSSM
parameter space is consistent with
successful searches for superpartners at LEP-II and at an upgraded Tevatron.
Since the results of these searches
will be known by the time the LHC begins its search for $\h$, we find
it useful to divide some of our results below into two
additional categories, according to
whether or not at least one superpartner will be detected by the time
the LHC begins its search for $\h$. Now, the ability of the Tevatron
to detect superpartners depends quite strongly on exactly which upgrade(s)
are implemented and on details of detector performance, whereas the reach of
LEP-II is essentially determined by kinematics. Therefore, for simplicity we
choose to use only LEP-II detection criteria, which we approximate by saying
that supersymmetry will be detected if a charged superpartner is lighter than
90 GeV. (Of course sneutrinos or non-LSP neutralinos may also be detected
at LEP-II, but we find that within our parameter space this correlates
extremely well with a charged sparticle also being accessible.)

Let us first consider the total production cross-section for $\h$ from gluon
fusion. In the narrow-width approximation this is related to the width of $\h$
into gluons by
$$
\sigma (pp\rightarrow \h) = {\pi^2 \over 8 \mhcubed} \Gamma (\h\rightarrow gg)
\>
\tau \int_\tau^1 {dx\over x} g(x,\mhsquared) g(\tau /x , \mhsquared)\> ,
\eqno(sigma)
$$
where $\tau = \mhsquared/s$ with ${\sqrt s} = 14 $ TeV. We work
consistently with $\alpha_s$ and
gluon distribution functions $g(x,\mhsquared)$
in leading order taken from [\cite{grv}].
In determining $\Gamma (\h\rightarrow gg)$, we
include all one-loop quark and squark loop graphs, with both stop and sbottom
mixing effects; the relevant formulas may be found for example in
[\cite{hhg}].
The top-quark loop is dominant for the standard model but we find that in the
constrained MSSM parameter space the effects of
stops and sbottoms can be quite significant, and can even result in
nearly complete destructive interference if a stop is lighter than about
100 GeV.

In Figure 1 we show the total production cross-section for each model
within our ensemble as a function
of $\mh$. Models which have a lightest squark (always a stop or sbottom)
with mass less than 200 GeV are denoted by an $\times$, and other models
are denoted by dots.
For comparison, the production cross-section for a standard model Higgs boson
of the same mass and with $m_{\rm top} = 175$ GeV (the standard model
cross-section varies only weakly with $m_{\rm top}$)
is shown as a solid line.
We should remark that for consistency we do
not  include higher order QCD corrections, because these have
only been computed for quark loops and not for squark loops (see [\cite{sdgz}]
and references therein).  Since these radiative corrections in the standard
model have been shown to be positive and large ($\sim$60\%), our results
should be regarded as conservative.
Full QCD corrections would of course have to
be included in any attempt to separate the
experimentally measured cross-section from the branching fraction
in an experimentally measured signal. In general,
we find that the sum of loops involving the squarks of the first two families
always makes a negligible contribution in the constrained parameter space. The
third family squark-loop contributions clearly
can either enhance or diminish the total cross-section.
Also, one should note that the squark-loop
effects are generally smaller if $\mh > 125$ GeV;
this is because larger $\mh$
requires large radiative corrections in the Higgs sector which in turn
corresponds to heavier stops and sbottoms.

In Figure 2 we show the total width (in keV) of $\h\rightarrow
\gamma \gamma$ as a function of $\mh$. The dominant contribution in the
standard model comes from $W$-boson loops; this is also true for
supersymmetric models when $\sin (\beta - \alpha) $
(the ratio of the coupling of $\h$ to electroweak
vector bosons to the corresponding coupling for a standard model Higgs boson)
is close to unity. This is always the case within the constrained super-unified
MSSM parameter space, except for some models with
$\tan\beta$ larger than 30 and light $A^0$.
The contributions to the amplitude coming from chargino loops are important for
chargino masses less than about 100 GeV,
and the sum of the third-family squark and slepton loop amplitudes can
also be significant. Even the sum of first and second-family sfermion
loops can change the total width by more than 10\%, if a slepton is within
the discovery reach of LEP-II. However, we find that complete destructive
interference between the $W$-loop and supersymmetric loops is never
possible within the constrained parameter space.
In Figure 2 we have used an $\times$ to denote
models in which the lightest charged supersymmetric particle is less than 90
GeV (and so may be detected at LEP-II), showing the susceptibility of the
$\h\rightarrow \gamma\gamma$ width to light superpartners.
Note that if supersymmetry is discovered at LEP-II, $\h$ cannot be heavier
than about 125 GeV with our allowed ranges of parameters in eq.~\(ranges).
(This upper bound would increase if $m_0 > 1$ TeV.)

To determine the total branching ratio of $\h\rightarrow \gamma\gamma$, we
must also account for all other decay modes. We compute the width
of $\h$ from tree-level decays into standard model fermion-antifermion
pairs using the formulas in e.g.~[\cite{hhg}], also including the significant
radiative corrections for the dominant $\h\rightarrow b \overline b$
decay mode as given in [\cite{gkls}].
We also include the decay widths of $\h$ into gluon pairs, and into off-shell
$WW^*$ and $ZZ^*$ states, which become increasingly important for larger
$\mh$ [\cite{km}].
Finally, we include decays of $\h$ into all kinematically allowed
neutralino, chargino, and sfermion-antisfermion pairs using the formulas in
[\cite{hhg}]. The possibility of these decay modes of $\h$ has
been neglected in many analyses,\footnote{$^\dagger$}{See, however,
refs.~[\cite{bddt},\cite{bbdkt},\cite{invisible}] which discuss the importance
of supersymmetric decays for $\h$ (and $H^0$ and $A^0$) in different
contexts.} but it is important to take them into account because they can
suppress the $\h\rightarrow \gamma\gamma$ branching fraction to an
unuseable level. In Figure 3 we show a scatterplot of the total width
of $\h$ into charginos and neutralinos as a function of $\mu$ for models
in which these decays are kinematically allowed. We find that within
the constrained parameter space the total
ino-ino decay widths of $\h$ can be as large as 50 MeV when
$\mu$ is positive (using the sign convention of [\cite{reviews},\cite{hhg}])
and as large as 10 MeV when $\mu$ is negative. On the other hand, even
if supersymmetry is not discovered at LEP-II (models denoted by $\times$'s),
the decay width of $\h$ into LSP pairs
can still be as large as 400 keV if $\mu > 0$ and 20 keV if $\mu < 0$.

When $\h$ decays into sneutrinos, charged sleptons, or
stops are kinematically allowed, we find that the corresponding widths are
usually even larger, typically tens or hundreds of MeV.
A subtlety arises if the Higgs mass is at or slightly below threshold of
a two body decay $\h\rightarrow AB$, since the simple 2-body decay formulas
[\cite{hhg}] will incorrectly yield a zero result for the calculated width.
The correct answer near or below threshold
is obtained only after calculating the full decay amplitude
with off-shell $A^*$ and $B^*$.  When $A$ and $B$ are superpartners
which eventually must each decay into a neutralino LSP ($\lsp$), the
result may be conveniently parameterized by
$$
\eqalignno{
\Gamma (\h \rightarrow A^* B^* \rightarrow X_A \lsp X_B \lsp) = &
\int_{\mlsp}^{\mh - \mlsp} {2 q_A^2 dq_A \over \pi}
\int_{\mlsp}^{\mh - q_A} {2 q_B^2  dq_B \over \pi}
\times &(fourbody) \cr
\Gamma (\h \rightarrow A^* B^*)
&{\Gamma (A^* \rightarrow X_A \lsp) \over
\left [(q_A^2 - m_A^2)^2 + \Gamma_A^2 m_A^2 \right ] }
{\Gamma (B^* \rightarrow X_B \lsp) \over \left [
(q_B^2 - m_B^2)^2 + \Gamma_B^2 m_B^2 \right ]} \> .
\cr }
$$
Here $\Gamma (\h \rightarrow A^* B^*)$ is the two-body decay width of
$\h$ into off-shell $A^*$ and $B^*$ of masses $q_A$ and $q_B$.
The decay widths of off-shell stops, charginos and neutralinos are always
very small if their masses are comparable to
$\mh/2$, so that off-shell decays of $\h$ into stops, charginos and
neutralinos should not be a concern. However, we do find models with
sneutrinos near the $\h \rightarrow \tilde\nu\overline{\tilde \nu}$ threshold,
and with $\Gamma({\tilde\nu}^* \rightarrow \nu \lsp )$ often exceeding 50 MeV.
In such cases we find that exactly at threshold ($\mh = 2 m_{\tilde \nu}$)
the contribution to the width of $\h$ is a few MeV, greatly reducing
the $\gamma \gamma$ signal.  The  contribution to the
$\h$ width drops quickly below
threshold, but it will still be in the hundred keV range
if $\mh$ is 3 GeV or in some cases even 4 GeV below the threshold.
The $\h$ width in these cases  is of course quite sensitive to how
near $2 m_{\tilde\nu}$ is to $\mh$, and to the sneutrino widths.
Similar statements apply if $\h$ decays to a
stau or right-handed selectron or smuon are within a few GeV of threshold.
These possibilities could be a concern if
evidence for sleptons is found at LEP-II.

Since standard model decays of the $\h$ typically have a combined width of
order 3 MeV, the $\h$ decays into supersymmetric states can
completely dominate the branching fractions.
While $\h$ might decay into potentially visible supersymmetric states,
or into invisible states [\cite{invisible}],
we find that the supersymmetric branching fraction of $\h$ can
exceed 15\% only if LEP-II detects supersymmetry.
More studies are needed to determine if visible supersymmetric decay
modes can be used to detect $\h$ (with the main background
presumably coming from continuum production of sparticles);
that will surely be difficult at the LHC, although for the heavier states
$H^0$ and $A^0$ it has been shown [\cite{bbdkt}]
to be a possibility for decays to pairs of second-lightest neutralinos.

Another factor which can severely diminish the $\h\rightarrow \gamma\gamma$
branching fraction is an enhanced coupling of $\h$ to $b\overline b$. In the
MSSM, this coupling is equal to the corresponding coupling of a standard
model Higgs boson multiplied by the factor
$ -\sin \alpha/\cos\beta$. This factor can be significantly
greater than 1 for all values of $\tan\beta$
if $A^0$ is not too heavy, which greatly enhances
$\Gamma (\h\rightarrow b\overline b) $ with a concomitant unfortunate
effect on the $\gamma\gamma$ branching fraction. As we have already
mentioned, a full treatment of radiative corrections in the Higgs sector
reveals that $-\sin\alpha /\cos\beta$ can also be
less than 1, reducing the $\h \rightarrow b\overline b$ branching
ratio and thus increasing the two-photon signal. However, this can only
happen if $\tan\beta > 30$.

All of these effects are included in Figure 4, which shows the
branching ratio Br($\h\rightarrow \gamma\gamma$) as a function of $\mh$
for models within the constrained parameter space, for
$\tan\beta < 30$ [Figure 4(a)] and $\tan\beta > 30$ [Figure 4(b)].
It is clear that
supersymmetric decay modes are less likely to be a problem if $\mh$
is near the higher end of its allowed range. This is because a heavier
$\h$ generally corresponds to large radiative corrections which in turn
are associated with a heavier supersymmetry-breaking scale, and
thus heavier sparticles.
For the same reason, supersymmetric contributions
to production and decay amplitudes also tend to be smaller for larger $\mh$.
The main ``risk factor" for the $\h\rightarrow \gamma\gamma$ signal
if $\mh > 125 $ GeV is therefore the
possibility of an enhanced $\h b\overline b$ coupling.
Note that if $\tan\beta > 30$, one can have Br($\h \rightarrow\gamma\gamma$)
as large as 1\%. Also it is important to note that $\tan\beta > 30$ implies
that $\mh > 100$ GeV, so that $\h$ cannot be discovered at LEP-II.

In Figure 5 we show the total cross-section times branching ratio
for the process $pp \rightarrow \> \h \rightarrow \gamma \gamma$
as a function of $\mh$ in the scenario that supersymmetry
cannot be discovered at LEP-II.
Even in this case there are models for which $\h$ decays into
LSPs are kinematically allowed, although they tend to be much less
important than in the scenario in which supersymmetry has already been
discovered before the LHC turns on, and they
are never responsible for killing the signal. Again
we have divided the models into small and large $\tan\beta$ categories. As we
remarked earlier, we do not include QCD radiative corrections to
the cross-section, which may increase it, since they have not been
fully computed for the supersymmetric case.
Since the SM QCD corrections increase the cross-section
by a factor of $\sim$1.6, QCD corrections to the supersymmetric
amplitudes could be similarly large.  Until these corrections
are fully understood, it will unfortunately not be possible to
separate $\sigma$ or Br($\h \rightarrow \gamma\gamma$) from the measured
$\sigma \times $ Br($\h \rightarrow \gamma\gamma$).
One would like to conclude from Fig.~5(a) that the leading order cross-section
times the branching fraction for
$pp\rightarrow\h\rightarrow\gamma\gamma$ will have to exceed
about 20 fb (before efficiencies) if supersymmetry
is not discovered at LEP-II and if $\tan\beta < 30$. This is indeed
true within the constrained super-unified MSSM framework, but as we shall
remark below, it need not hold in a more general model framework.

Figure 6 shows the total cross-section times branching ratio for models
in the complementary scenario in which
at least one supersymmetric particle should be detected at LEP-II.
In this case, the $\h\rightarrow\gamma\gamma$
signal may be endangered not only by the possibility of competing
supersymmetric decay modes (models denoted by $\times$'s) but by an enhanced
width into $b\overline b$ or by sfermion loop effects which can decrease
the total $\h$ production rate and the $\gamma\gamma$ width.

Finally, in
Figure 7 we show the dependence of the cross-section times branching ratio
on $m_{A^0}$, for models with $\mh> 95$ GeV so that $\h$
cannot be discovered at LEP-II. Note that even when ${A^0}$ is
very heavy, there are models in which $\h$ cannot be discovered at the LHC,
primarily but not exclusively because of the possibility of
supersymmetric decay modes.

Let us now turn to the question of how dependent our results are
on the choice of boundary conditions for the soft terms.
One particularly well-motivated generalization
of the universal soft-breaking boundary condition is to add D-term
contributions [\cite{dterms},\cite{kolda}]
to the scalar squared masses, since these will
naturally maintain the cancellation of flavor-changing neutral currents.
Such contributions
should arise whenever the unbroken gauge group at high energies
has rank $>4$, and could easily be comparable in magnitude to the
usual soft-breaking parameters even if the scale at which the gauge group
breaks down to $SU(3)_C \times SU(2)_L \times U(1)_Y$ is arbitrarily high.
In order to check the generality of our results,
we have generated an ensemble
of models as discussed above, but also including D-term contributions to
scalar masses that could arise from the spontaneous breaking of
an arbitrary subgroup of $E_6$,
as parameterized in [\cite{kolda}]. We find that the most important
change induced by these generalized boundary conditions in the
results outlined above is that $A^0$ can be lighter,
corresponding to a larger $\h b\overline b$ coupling.
This difference is particularly significant
if supersymmetry is not discovered at LEP-II and $\tan\beta < 30$;
with universal scalar mass boundary conditions one then
finds $m_{A^0} > 200 $ GeV throughout the constrained parameter space,
but with large D-term contributions to scalar masses one can find models
with $m_{A^0}$ as low as 110 GeV.  The effect of this is to substantially
reduce the $\h\rightarrow \gamma\gamma$ branching fraction by increasing
the $\h\rightarrow b\overline b$ width. This may occur if
$ 4D_X -  D_S + 3 D_{\hat Y}> 0$ in the notation of [\cite{kolda}],
corresponding to a change in the scalar mass boundary conditions with
$\Delta (m^2_{H_u} - m^2_{H_d} ) > 0$.
As in the case of universal scalar mass boundary conditions, we find that
the $\h\rightarrow\gamma\gamma$
signal cannot be enhanced by a smaller $\h b\overline b$ coupling
unless $\tan\beta > 30$. While the leading order cross-section times
the branching fraction before efficiencies
is always larger than about 20 fb if supersymmetry is
not discovered at LEP-II and $\tan\beta < 30$ in the case of
universal scalar mass boundary conditions [see Fig.~5(a)],  this
need not hold if D-term contributions to scalar masses are large
compared to the universal soft supersymmetry-breaking masses and
have the appropriate signs.

In this paper we have examined the variation of the $pp \rightarrow \h
\rightarrow \gamma\gamma$ signal at the LHC within the framework of the
constrained MSSM. Certainly if LHC does not detect $\h$ it will
not be possible to conclude that $\h$ does not exist.
We note that prospects for discovering $\h$ through the two-photon decay mode
depend strongly on whether or not supersymmetry will have been discovered
at LEP-II (or an upgraded Tevatron). If supersymmetry is not discovered
at LEP-II, then $\h \rightarrow $ SUSY decays may still be allowed, but
will be sufficiently kinematically disfavored that they cannot reduce the
$\h \rightarrow \gamma\gamma$ signal by more than about 15\%.
In this scenario, the most important risk factor for discovering
$\h$ at the LHC is the possibility that
$- \sin \alpha /\cos \beta $ is significantly
greater than 1. The $\h\rightarrow\gamma\gamma$ signal could also be
dramatically enhanced by $-\sin\alpha /\cos\beta < 1$ so that
Br($\h\rightarrow\gamma\gamma$) can be as large as 1\%,
but we find that this requires
$\tan\beta > 30$ within the constrained MSSM parameter space.
If one or more supersymmetric particles are detected by the time the LHC
begins its search for $\h$, careful study will be required to determine
if the $\h \rightarrow \gamma\gamma$ signal at the LHC is still viable.

\noindent Acknowledgments:
We are grateful to Martin Beneke and Chris Kolda for valuable discussions.
G.L.K. appreciates hospitality at the Institute for
Theoretical Physics in Santa Barbara, where he was visiting when this
work was completed.
J.D.W. would like to acknowledge the hospitality of the SLAC theory group and
the BNL high energy theory group.
This work was supported in part by the U.S. Department of Energy.

\references

\refis{reviews}
For reviews, see H.~E.~Haber and G.~L.~Kane, \prpts 117, 75, 1985;
H.~E.~Haber, ``Introductory low-energy supersymmetry",
TASI-92 lectures, hep-ph/9306207.

\refis{hhg} J.~F.~Gunion, H.~E.~Haber, G.~L.~Kane, and S.~Dawson,
``The Higgs Hunter's Guide", Addison-Wesley 1991; errata hep-ph/9302272.

\refis{kkw} G.~L.~Kane, C.~Kolda, and J.~Wells;
\prl 70, 2686, 1993; J.~Espinosa and M.~Quiros, \pl B302, 51, 1993.

\refis{kkrw} G.~L.~Kane, C.~Kolda, L.~Roszkowski, and J.~Wells,
\pr D49, 6173, 1994.

\refis{ehsv} R.~K.~Ellis, I.~Hinchliffe, M.~Soldate, and J.~J.~van der Bij,
\np B297, 221, 1988.

\refis{gkw} J.~F.~Gunion, G.~Kane, and J.~Wudka, \np B299, 221, 1988.

\refis{ggn} J.~F.~Gunion, G.~Gamberini, and S.~F.~Novaes,
\pr D38, 3481, 1988.

\refis{bbsp} V.~Barger, M.~S.~Berger, A.~L.~Stange, and R.~J.~N.~Phillips,
\pr D45, 4128, 1992.

\refis{bbkt92} H.~Baer, M.~Bisset, C.~Kao, and X.~Tata,
\pr D46, 1067, 1992.

\refis{kz} Z.~Kunszt and F.~Zwirner, \np B385, 3, 1992.

\refis{bbdkt} H.~Baer, M.~Bisset, D.~Dicus, C.~Kao, and X.~Tata,
\pr D47, 1062, 1993; H.~Baer, M.~Bisset, C.~Kao, and X.~Tata,
\pr D50, 316, 1994.

\refis{cms} CMS collaboration, Technical Proposal, Report CERN-LHCC 94-38.

\refis{atlas} ATLAS collaboration, Technical Proposal, Report CERN-LHCC 94-43.

\refis{sdgz} A.~Djouadi, M.~Spira, and P.~M.~Zerwas, \pl B264, 440, 1991;
M.~Spira, A.~Djouadi, D.~Graudenz, and P.~M.~Zerwas,
``Higgs boson production at the LHC", hep-ph/9504378.

\refis{kop} B.~Kileng, \journal Z.~Phys., C63, 87, 1994;
B.~Kileng, P.~Osland, and P.~N.~Pandita, ``Production and
two-photon decay of the MSSM scalar Higgs bosons at the LHC",
hep-ph/9506455.

\refis{konig} H.~K\"onig, ``The decay $H_2^0 \rightarrow gg$ within the
minimal supersymmetric standard model", hep-ph/9504433.

\refis{grv} M.~Gl\"uck, E.~Reya, and A.~Vogt,
\journal Z.~Phys., C53, 651, 1992.

\refis{gkls} S.~G.~Gorishny, A.~L.~Kataev, S.~A.~Larin, and
L.~R.~Surgaladze, \journal Mod. Phys. Lett., A5, 2703, 1990.

\refis{km} W.-Y.~Keung and W.~J.~Marciano, \pr D30, 248, 1984.

\refis{finetuning} R.~Barbieri and G.~F.~Giudice, \np B306, 320, 1988;
B.~de Carlos and J.~A.~Casas, \pl B309, 320, 1993;
G.~W.~Anderson and D.~J.~Casta\~no, \pl B347, 300, 1995 and
hep-ph/9412322.

\refis{pdg} Particle Data Group, L.~Montanet et. al. \pr D50, 1173, 1994.

\refis{bsgamma} CLEO Collaboration, \prl 74, 2885, 1994.

\refis{bddt} H.~Baer, D.~Dicus, M.~Drees, and X.~Tata,
\pr D36, 1363, 1987.

\refis{invisible}  K.~Griest and H.~Haber, \pr D37, 719, 1988;
 J.~Gunion, \prl 72, 199, 1994; S.~G.~Frederiksen, N.~Johnson,
G.~L.~Kane, and J.~Reid, \pr D50, R4244, 1995.

\refis{backgrounds} See D.~Dicus and S.~Willenbrock, \pr D37, 1801, 1988
and references therein.

\refis{fnal} S.~Mrenna and G.~L.~Kane, hep-ph/9406337. See also
W.~Marciano, A.~Stange, and S.~Willenbrock \pr D50, 4491, 1994 and
J.~Gunion and T.~Han, \pr D51, 1051, 1995.

\refis{effpot} J.~Ellis, G.~Ridolfi, and F.~Zwirner, \pl B262, 477, 1991;
M.~Drees and M.~M.~Nojiri, \pr D45, 2482, 1992.

\refis{sugra} A.~Chamseddine, R.~Arnowitt, and P.~Nath,
\prl 49, 970, 1982; H.~P.~Nilles, \pl 115B, 193, 1982;
L.~E.~Ib\'a\~nez, \pl 118B, 73, 1982;
R.~Barbieri, S.~Ferrara and C.~Savoy, \pl 119B, 343, 1982;
L.~Hall, J.~Lykken and S.~Weinberg, \pr D27, 2359, 1983;
P.~Nath, R.~Arnowitt and A.~Chamseddine, \np B227, 121, 1983.

\refis{clm} J.~A.~Casas, A.~Lleyda, and C.~Mu\~noz, ``Strong constraints
on the parameter space of the MSSM from charge and color breaking minima",
preprint FTUAM 95/11, IEM-FT-100/95, hep-ph/9507294.

\refis{dn} M.~Drees and M.~M.~Nojiri, \pr D47, 376, 1993;
M.~Srednicki, R.~Watkins, and K.~A.~Olive, \np B310, 693, 1988.

\refis{dterms} M.~Drees, \pl B181, 279, 1986; J.~S.~Hagelin and
S.~Kelley, {\it Nucl.~Phys.}~B342, 1990;
Y.~Kawamura, H.~Murayama, and M.~Yamaguchi,
\pl B324, 52, 1994, \pr D51, 1337, 1995; Y.~Kawamura and M.~Tanaka,
{\it Prog.~Theor.~Phys.} {\bf 91}, 949, 1994.

\refis{kolda} C.~Kolda and S.~P.~Martin, ``Low-energy supersymmetry
with D-term contributions to scalar masses", hep-ph/9503445.

\endreferences
\oneandahalfspace
\subhead{Figure Captions}

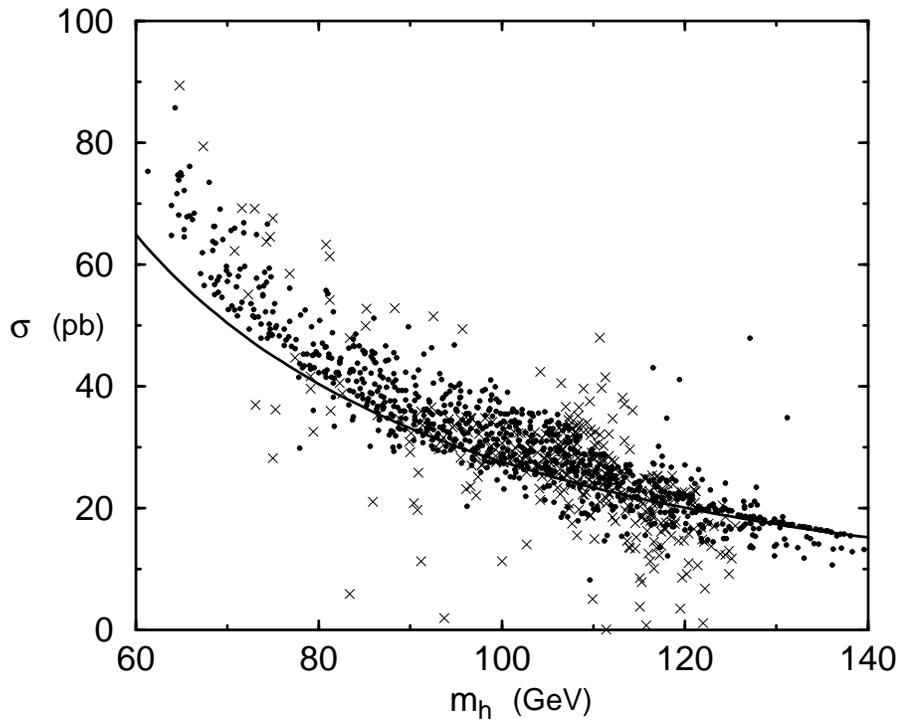
\noindent{} Figure 1: The total production cross-section (in pb)
for $\h$ in pp collisions at $\sqrt{s} = 14$ TeV as a function of $\mh$.
Each point
represents a different set of model parameters which satisfy all experimental
and theoretical
constraints. Models for which the mass of the
lightest squark (usually a stop) is less than 200 GeV are represented by an
$\times$,
while all other models are represented by a dot. As a convenient reference,
we also show as a solid line the total production cross-section
of a standard model Higgs
boson with the same mass and $m_{\rm top} = 175$ GeV.
Higher order QCD corrections are not included since they have not been fully
computed for the supersymmetric case (see comments in text).

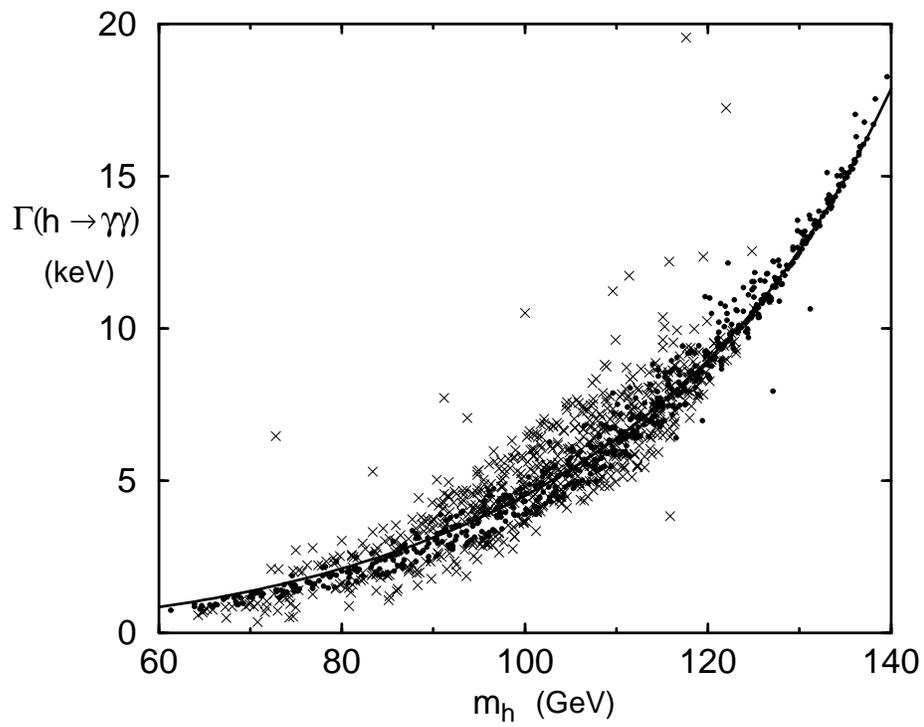
\noindent{} Figure 2: The total width (in keV)
for $\h$ decays to two photons as a function of $\mh$.
Models for which the lightest charged supersymmetric particle
is less than 90 GeV (and thus detectable at LEP-II) are represented by an
$\times$,
while all other models are represented by a dot.
We also show as a solid line the same decay width for a standard model
Higgs boson of the same mass and $m_{\rm top} = 175$ GeV.

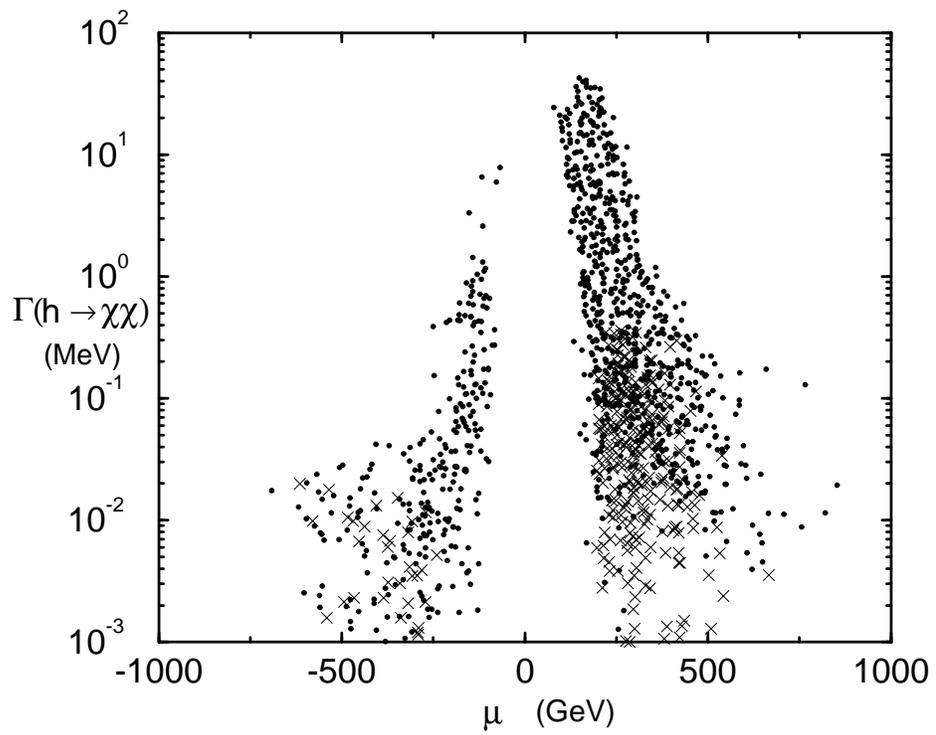
\noindent{} Figure 3: The total width (in MeV)
for $\h$ decays into neutralino and chargino pairs,
as a function of $\mu$ (using the sign convention of
[\cite{reviews},\cite{hhg}]).
Models for which a supersymmetric particle should be detected at LEP-II
are denoted by a dot, and other models in which $\h$ can decay to LSP
pairs are denoted by an $\times$.

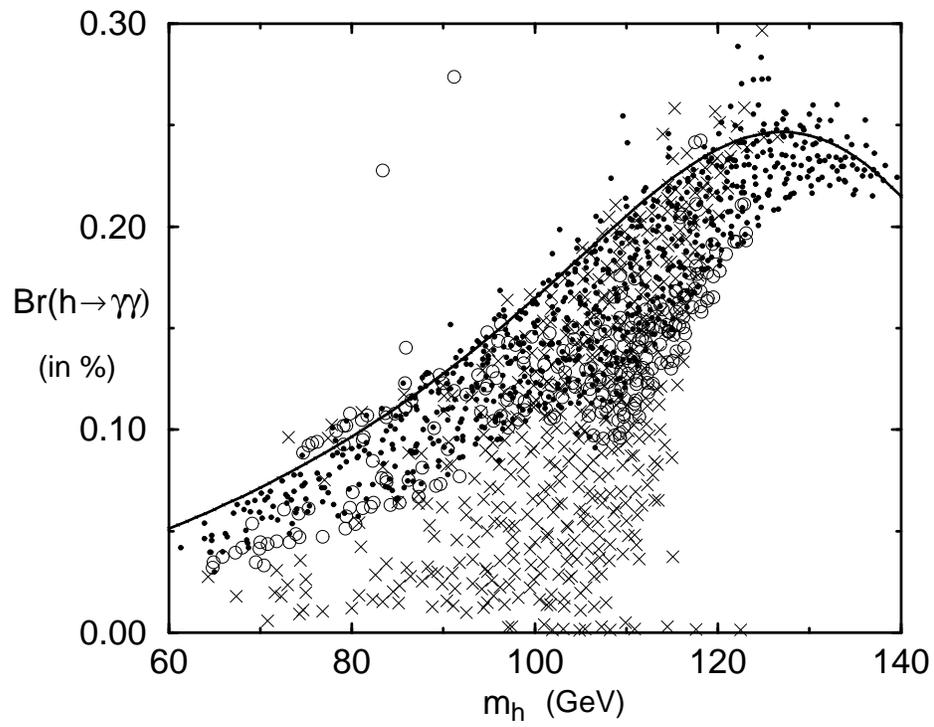
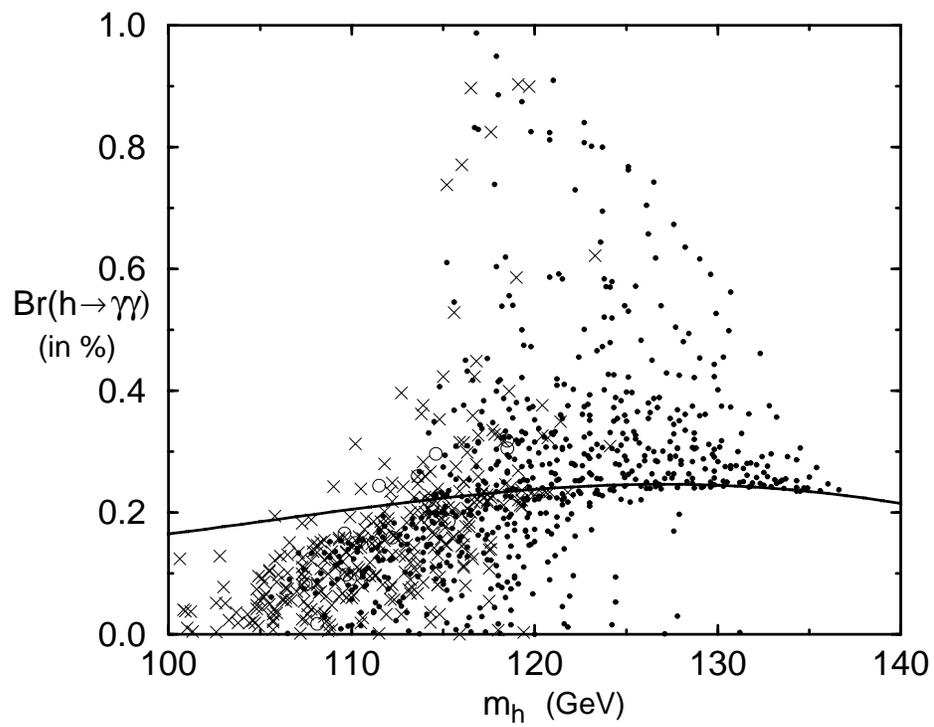
\noindent{} Figure 4: The branching fraction Br($\h\rightarrow \gamma\gamma)$
(in \%) as a function of $\mh$, for models with $\tan\beta < 30$ (a)
and $\tan\beta > 30$ (b).
Models for which $\h$ is kinematically allowed to decay into
pairs of supersymmetric particles are represented by an $\times$, and
models
for which supersymmetry is likely to be discovered at LEP-II
(but for which supersymmetric decays of $\h$ are not kinematically allowed)
are represented by an open circle, while all remaining models are
denoted by a dot.
The solid line is the same branching fraction for a standard model
Higgs boson of the same mass and with $m_{\rm top} = 175$ GeV.

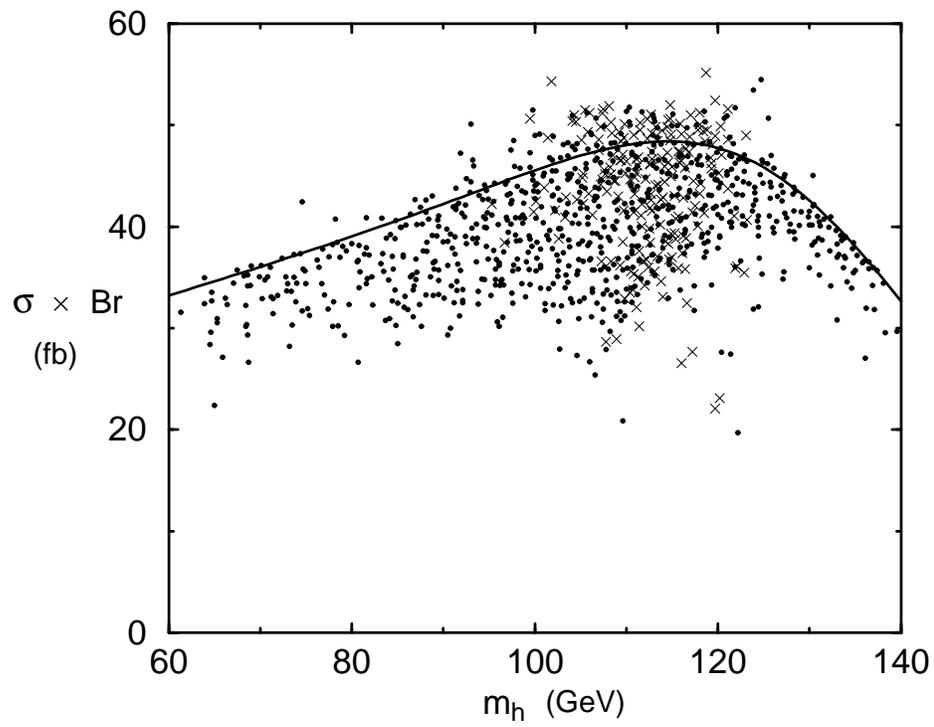
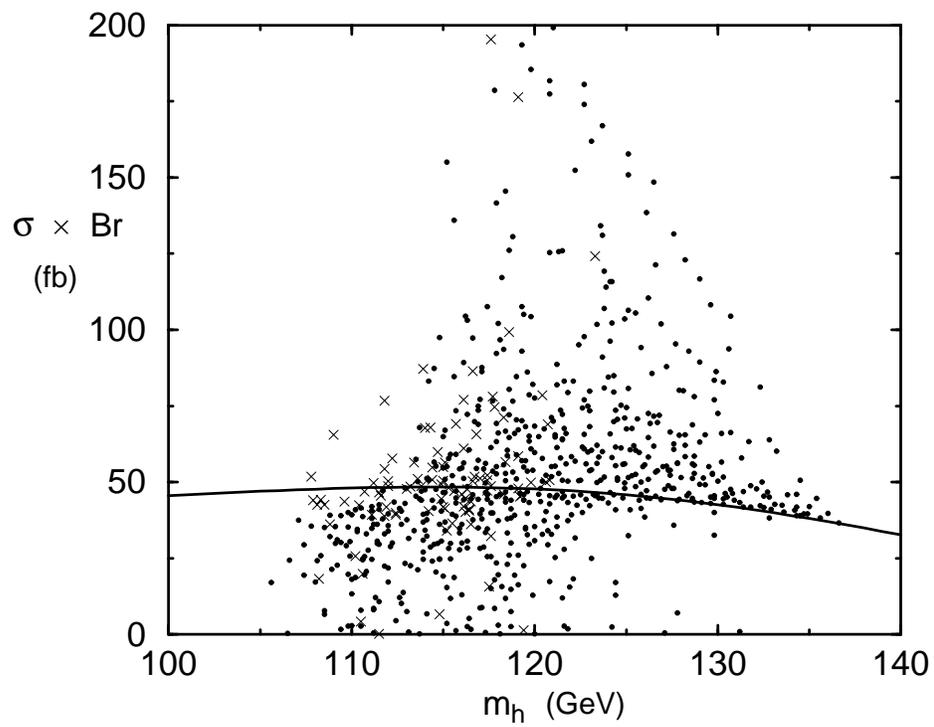
\noindent{} Figure 5: The cross-section times branching fraction
(in fb) for $pp \rightarrow \h \rightarrow \gamma\gamma$ as a function
of $\mh$, for models in which supersymmetry cannot be discovered
at LEP-II, with $\tan\beta < 30$ (a) and $\tan\beta > 30$ (b).
Models for which $\h$ is kinematically allowed to decay into
pairs of supersymmetric particles are represented by an $\times$.
No efficiencies are included, and again we emphasize that
QCD corrections to the
production cross-section are not included.
The solid line is the same cross-section times branching fraction
for a standard model
Higgs boson of the same mass and $m_{\rm top} = 175$ GeV.

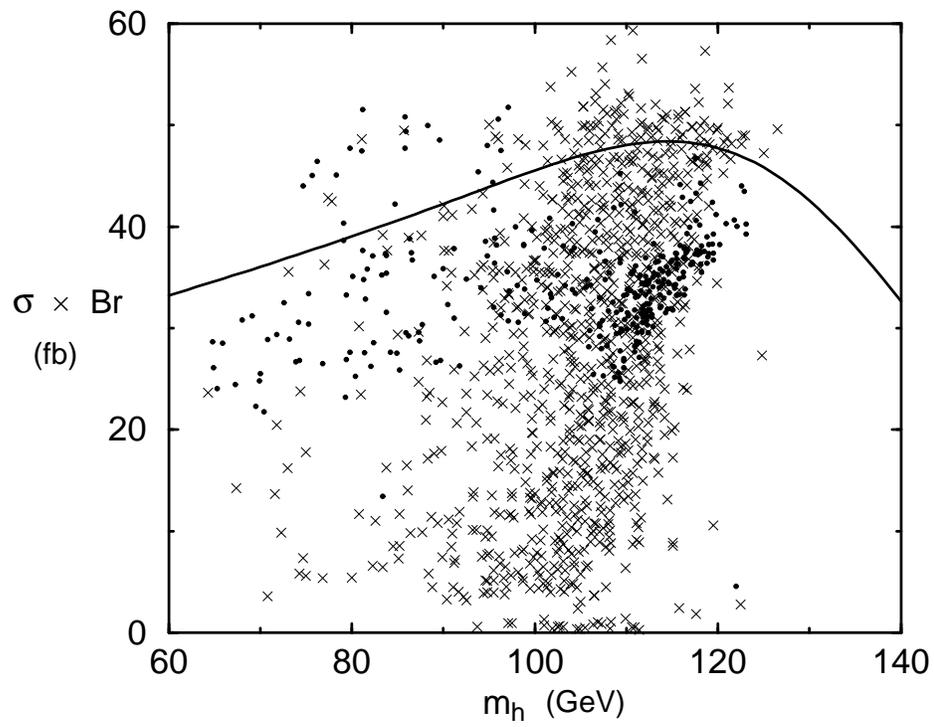
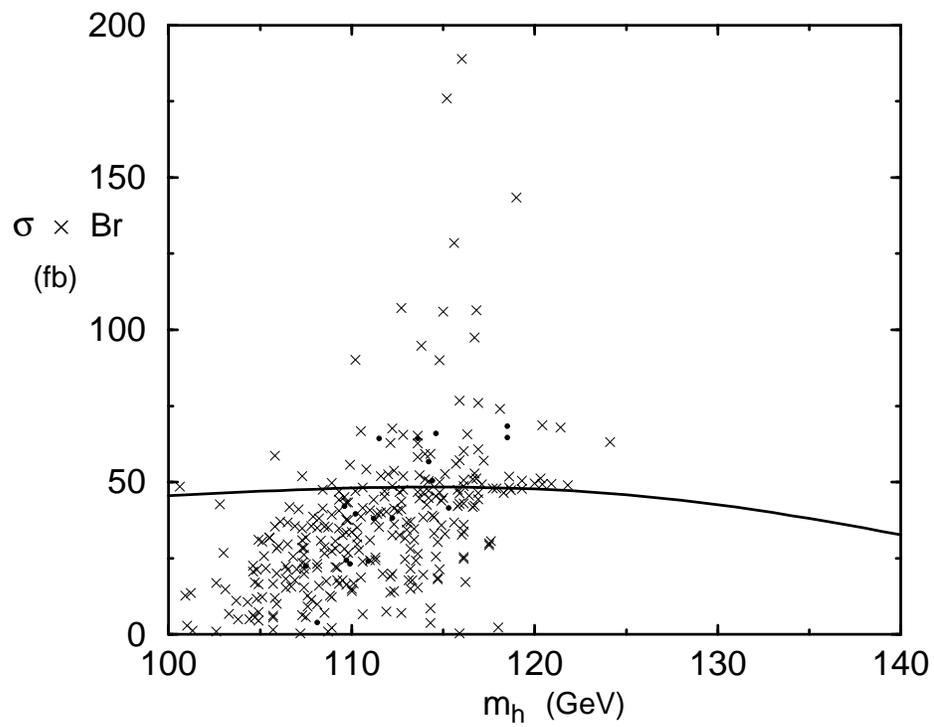
\noindent{} Figure 6: The same as Figure 5, but for models
in which supersymmetric particles can be detected at LEP-II.

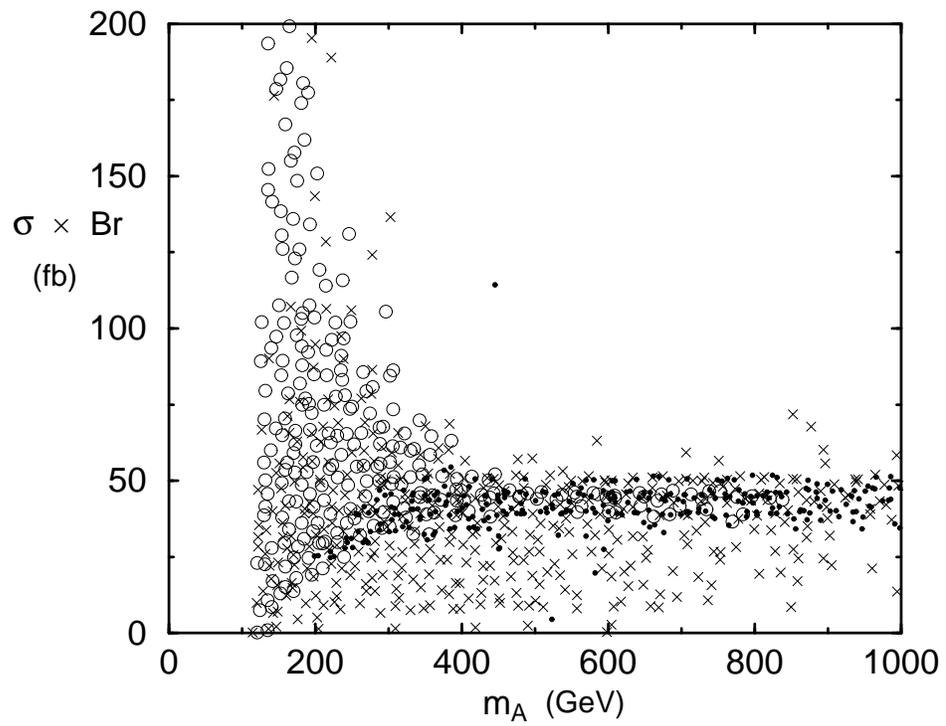
\noindent{} Figure 7: The cross-section times branching fraction for
$pp \rightarrow \h \rightarrow \gamma\gamma$
(in fb) as a function of $m_{A^0}$, for models with $\mh > 95$ GeV, so that
$\h$ cannot be discovered at LEP-II. Models in which $\h$ has kinematically
allowed supersymmetric decays are denoted by an $\times$, and the remaining
models
are divided into $\tan\beta > 30$ (open circles) and $\tan\beta < 30$ (dots).

\endit\end